\documentclass[%
superscriptaddress,
 amsmath,amssymb,
onecolumn
]{revtex4-2}
\usepackage{setspace}

\usepackage{graphicx}% Include figure files
\usepackage{dcolumn}% Align table columns on decimal point
\usepackage{booktabs}
\usepackage[dvipsnames]{xcolor}
\usepackage{color}
\usepackage{tikz}
\usetikzlibrary{shapes}
\usepackage{soul}
\usepackage{hhline}

\usepackage{mathtools,mathbbol}
\usepackage[bookmarks,colorlinks,citecolor=blue,linkcolor=blue]{hyperref}
\usepackage{bm}% bold math
\newcommand{\pe}{\mathrm{Pe}}
\newcommand{\pec}{\mathrm{Pe}_c}
\newcommand{\peg}{\mathrm{Pe}_g}
\newcommand{\pecg}{\mathrm{Pe}_{cg}}

\begin{document}
\setstretch{1.3}

\newcommand{\markerone}{\raisebox{0.6pt}{\tikz{\node[draw,scale=0.5,regular polygon, regular polygon sides=4,fill=black!100!](){};}}}
\newcommand{\markertwo}{\raisebox{0.6pt}{\tikz{\node[draw,scale=0.3,regular polygon, regular polygon sides=3,fill=black!100!](){};}}}
\newcommand{\markerthree}{\raisebox{0.6pt}{\tikz{\node[draw,scale=0.3,regular polygon, regular polygon sides=3,fill=black!100!,rotate=90](){};}}}
\newcommand{\markerfour}{\raisebox{0.6pt}{\tikz{\node[draw,scale=0.3,regular polygon, regular polygon sides=3,fill=black!100!,rotate=270](){};}}}
\newcommand{\markerfive}{\raisebox{0.6pt}{\tikz{\node[draw,scale=0.3,regular polygon, regular polygon sides=3,fill=black!100!,rotate=180](){};}}}
\newcommand{\markersix}{\raisebox{0.6pt}{\tikz{\node[draw,scale=0.5,circle,color=red!80,fill=red!80!](){};}}}
\newcommand{\markerzero}{\raisebox{0.6pt}{\tikz{\node[draw,scale=0.5,circle,color=OliveGreen!80,fill=OliveGreen!80!](){};}}}
\newcommand{\markerseven}{\raisebox{0.6pt}{\tikz{\node[draw,scale=0.5,regular polygon, regular polygon sides=5,fill=black!100!](){};}}}
\newcommand{\markereight}{\raisebox{0.6pt}{\tikz{\node[draw,scale=0.4,star, ultra thin,fill=black!100!](){};}}}
\newcommand{\markernine}{\raisebox{0.6pt}{\tikz{\node[draw,scale=0.5,regular polygon, regular polygon sides=6,fill=black!100!](){};}}}
\newcommand{\markerten}{\raisebox{0.6pt}{\tikz{\node[draw,scale=0.4,diamond,fill=black!100!](){};}}}
\newcommand{\markereleven}{\raisebox{0.6pt}{\tikz{\node[draw,scale=0.5,cross out, ultra thick,fill=black!100!](){};}}}
\newcommand{\markertwelve}{\raisebox{0.6pt}{\tikz{\node[draw,scale=0.4,cross out, ultra thick,fill=black!100!,rotate=45](){};}}}
\newcommand{\markerthirteen}{\raisebox{0.6pt}{\tikz{\node[draw=red!100!,scale=0.5,circle,fill=red!100!](){};}}}
\newcommand{\markerfourteen}{\raisebox{0.6pt}{\tikz{\node[draw=green!70!blue,scale=0.5,regular polygon, regular polygon sides=4,fill=green!70!blue](){};}}}
\newcommand{\markerfifteen}{\raisebox{0.6pt}{\tikz{\node[draw=orange!100!,scale=0.3,regular polygon, regular polygon sides=3,fill=orange!100!,rotate=0](){};}}}
\newcommand{\markersixteen}{\raisebox{0.6pt}{\tikz{\node[draw=cyan!50!blue,scale=0.4,diamond,fill=cyan!50!blue](){};}}}
\newcommand{\markereighteen}{\raisebox{0.6pt}{\tikz{\node[draw=pink!50!magenta,scale=0.3,regular polygon, regular polygon sides=3,fill=pink!50!magenta,rotate=270](){};}}}

\title{Rheology of bidisperse suspensions at the colloidal-to-granular transition}

\author{Xuan Li}
\affiliation{School of Engineering, The University of Edinburgh, Edinburgh EH9 3JL, United Kingdom}

\author{John R. Royer}%
\affiliation{School of Physics and Astronomy, The University of Edinburgh, Edinburgh EH9 3FD, United Kingdom}

\author{Christopher Ness}%
\affiliation{School of Engineering, The University of Edinburgh, Edinburgh EH9 3JL, United Kingdom}

\date{\today}

\begin{abstract}
We use particle-based simulation to study the rheology of
dense suspensions comprising
mixtures of small colloids and larger grains,
which exhibit shear thinning at low shear rates and shear thickening at high shear rates.
By systematically varying the volume fraction of the two species, we demonstrate a monotonic increase in viscosity when grains are added to colloids,
but, conversely, a nonmonotonic response
in both the viscosity and shear thickening onset when colloids are added to grains.
Both effects are most prominent at intermediate shear rates where diffusion and convection play similar roles in the dynamics.
We rationalise these results by measuring the maximum flowable volume fraction as functions of the
P\'eclet number and composition,
showing that in extreme cases increasing the solids content can allow a jammed suspension to flow.
These results establish a constitutive description for the rheology of bidisperse suspensions across the colloidal–to-granular transition,
with implications for flow prediction and control in multicomponent particulate systems.
\end{abstract}

\maketitle

\section{Introduction}
Suspensions containing mixed Brownian colloids and larger, non-Brownian grains are prevalent across industry,
with important examples being cement~\cite{aitcin2000cements} and slurries~\cite{chen2009rheological}.
%, and ice cream~\cite{goff1997colloidal}.
When prepared with moderate to high solids volume fraction~$\phi$,
which is often desirable for function,
these materials exhibit complex rheological behaviour including shear thinning, thickening and, as $\phi$ approaches its maximal value $\phi_m$, flow arrest~\cite{de1985hard,foss2000structure,brown2014shear,ness2022physics}.
In fully non-Brownian systems with bidisperse grains, the suspension viscosity diverges as a power law at large $\phi$,
with~\cite{ikeda2012unified, pednekar2018bidisperse} 
\begin{equation} 
\eta(\phi) \sim (1-\phi / \phi_m)^{-\beta}.
\label{eq:visc_scale}
\end{equation}
The jamming point $\phi_m$ depends on the particle friction coefficient $\mu$ and is also a nonmonotonic function of the nominal size ratio $\Delta$ and volumetric mixing ratio $\alpha$ of the two species (see e.g.~\cite{yu1991estimation,anzivino2023estimating,shapiro1992random}).
While geometric packing models are sufficient to predict the changes in constitutive curves observed when small particles are added to a system of larger ones~\cite{singh2024rheology,farris1968prediction},
they neglect the additional complexity present when the smaller species is colloidal.
Understanding the physics of such systems is likely crucial for characterizing phenomenology such as superplasticization reported in concrete formulation and elsewhere~\cite{roussel2010steady,van2018concrete,flatt2004towards,ito2019shear,thievenaz2021droplet}.
Moreover,
experimental measurements in bidisperse colloidal and granular suspensions \cite{cwalina2016rheology,madraki2017enhancing,madraki2018transition} report trends similar to those we explore here, highlighting the practical importance of understanding the interplay of Brownian motion, particle size, and rheology.

In mixed colloidal and granular systems, taking $\alpha$ as the fraction of the solid volume occupied by grains,
the timescale introduced by thermal motion 
${6 \pi \eta_0 a^3}/{k_B T}$
 leads to rate-dependence in $\phi_m$
 and influences the functional form of $\eta(\phi)$.
Here $\eta_0$ is the solvent viscosity,
$a$ is a characteristic particle size and $k_BT$ is the thermal energy.
Crucially,
changing $\alpha$ even at fixed shear rate $\dot{\gamma}$
not only affects the geometry as in non-Brownian bidisperse suspensions
but also the relative importance of 
diffusion,
through the $\alpha$-dependence of the characteristic particle size used when calculating the P\'eclet number 
$\pe = {6 \pi \eta \dot\gamma a^3}/{k_B T}$.
For (nearly) monodisperse spheres at very large $\pe$ the particles are effectively non-Brownian grains,
so that the relevant $\phi_m$ is the frictional random loose packing limit (e.g. $\phi_m^{\mu=1}\approx0.57$).
At $\pe\approx 1$,
recent work shows that thermal motion releases frictional constraints so that the relevant $\phi_m$ becomes frictionless random close packing $\phi_m^{\mu=0}\approx 0.64$~\cite{li2024simulating}.
%Here $\mu$ is the particle-particle friction coefficient.
Meanwhile in the limit of vanishing $\pe$
the viscosity diverges at the glass transition $\phi_G\approx 0.58$ with differing functional form
 $\eta(\phi)\sim\exp{(k/(\phi_G-\phi))}$~\cite{hunter2012physics,ikeda2012unified}.
In characterizing flowing colloidal-granular suspensions it is thus crucial to understand the interplay between the frictionless and frictional character as functions of the flow rate and composition~\cite{guy2015towards,guy2018constraint}.

Real-world suspensions have, in practice, continuous
mono- or multimodal particle size distributions often spanning orders of magnitude in $a$ from nano- to millimetric~\cite{
tari1999influence,
vlasak2011effect,
bauer2014rheological,
mehdipour2017effect,
hlobil2022surface},
obscuring the basic physics that link composition to constitutive behavior.
To make this problem tractable we study by particle-based simulation a model bidisperse colloidal($c$)-granular($g$) system of frictional spheres with size ratio $\Delta\equiv a_g/a_c=5$,
chosen to ensure the separation in their diffusive timescales $\sim\Delta^3$ is large.
The particles experience the same set of governing forces,
so the system now has two separate
dimensionless shear rates (P\'eclet numbers),
with $\pec={6 \pi \eta \dot\gamma a_c^3}/{k_B T}$ for the colloids and likewise $\peg={6 \pi \eta \dot\gamma a_g^3}/{k_B T}$ for the grains.
Varying $\alpha$ varying from $0$ to $1$ over a wide range of dimensionless shear rates, our model allows us to map the full rheology of this bidisperse system
at $\phi$ close to the varying $\phi_m$.

In this article, we do three things:
(i) we present the first systematic simulation study of the rheology of a colloid-grain mixture, showing separately the impact of adding colloids (grains) to a granular (colloidal) suspension. The rheology shows characteristic shear thinning followed by thickening with increasing $\pe$, with the onset and extent of thickening being sensitive to $\alpha$;
(ii) we provide the first simulation demonstration that adding colloids to a granular suspension can enhance flowability even as $\phi$ increases. This can manifest as a modest viscosity reduction, or even as unjamming of a granular packing by adding a small volume of colloids;
(iii) we explain the above phenomenology by mapping   $\phi_m$ as a function of both $\alpha$ and a suitably defined dimensionless shear rate.
Collectively, these results offer a rationale for predicting the constitutive behavior of suspensions of mixed colloids and grains,
and provide a starting point for addressing the rheology of more complex multi-species suspensions, a major challenge in geophysics, formulation science and other application areas.

\section{Simulation method overview}
We simulate the trajectories of spheres with motion governed by Langevin equations comprising three pairwise force and torque contributions: direct contacts, short-ranged hydrodynamics and Brownian noise.
Contact forces are modeled as linear springs with a tangential component regulated by a friction coefficient $\mu=1$.
Following standard convention for dense suspensions, hydrodynamics enter through single-body drag and pairwise lubrication~\cite{ball1997simulation,seto2013discontinuous,cheal2018rheology}.
Brownian forces and torques are generated on each time step to satisfy fluctuation–dissipation theorem as described in~\cite{li2024simulating}.
Our model thus incorporates sufficient microscopic physics to capture the suspension rheology around the colloidal-to-granular crossover as a function of $\mathrm{Pe}$.
A summary of the governing equations and parameter values is given in the Appendix,
and a comprehensive model rationale and description is given in~\cite{li2024simulating}.

The model is implemented in LAMMPS,
with shear applied by deforming the
simulation box at fixed volume $V$, 
updating the particle displacements following the Velocity-Verlet algorithm.
Simulations are initialized in randomized non-overlapping configurations of up to $10^4$ particles,
sufficient to mitigate finite-size effects.
To prevent crystallization, our nominally bidisperse system
comprises particles with size ratio 1:1.4:5:7,
with colloid and grain volumes given by $V_c=V_1 + V_{1.4}$ and $V_g=V_5 + V_{7}$. The total solids volume fraction is $\phi=(V_g+V_c)/V$, with the volumetric mixing ratio $\alpha=V_g/(V_c + V_g)$
and $V_l=V-V_c-V_g$ the free `liquid' volume.
Snapshots in Fig.~\ref{fig1}(a) show the simulated system at a range of $\alpha$.
All particles experience the same forces, including the Brownian noise, but
we refer to the smaller ones as colloids since their diffusive
timescale is $125\times$ shorter than that of the larger grains.
We subject the system to constant $\dot\gamma$ for sufficiently long run time $t$ so that the shear stress $\Sigma_{xy}$
reaches a steady-state at large strains $\dot{\gamma}t$,
before taking the suspension viscosity as $\eta = \Sigma_{xy}/ \eta_0 \dot\gamma$. To obtain constitutive curves, we adjust both $\dot{\gamma}$ and $k_BT$ to cover a wide range of $\mathrm{Pe}_c=10^{-2}-10^6$ while maintaining both near inertia-free and hard-sphere conditions (see Appendix).

\section{RESULTS: Mixing colloids and grains}

%===========================
%===========================
%===========================
%===========================
%===========================
\begin{figure*}
        \includegraphics[trim = 0cm 20mm 0mm 0cm, clip,width=0.985\textwidth,page={1}]{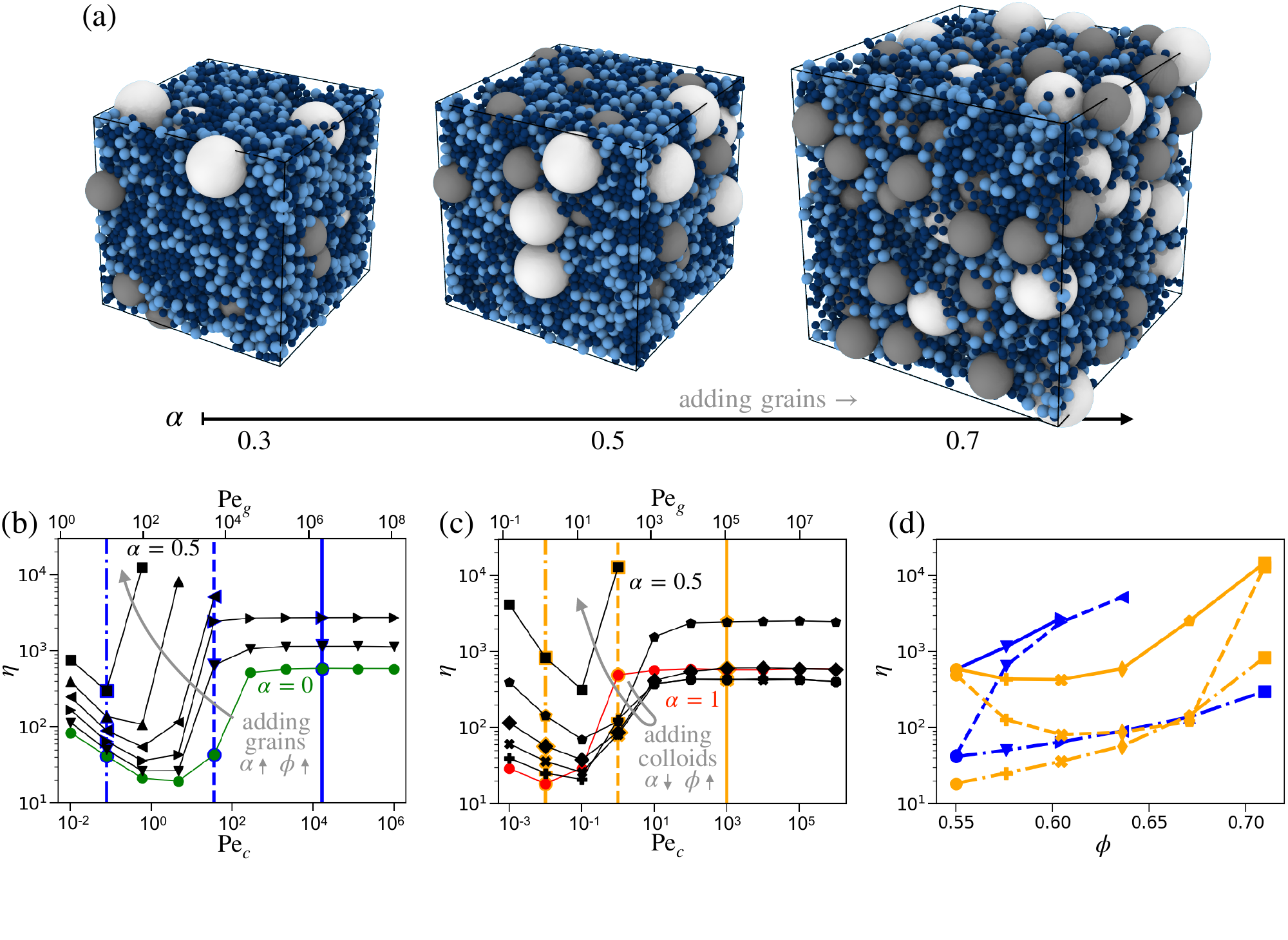}
        \caption{
        Rheological impact of adding grains and colloids.
        (a) Snapshots of the simulated system at a range of compositions $\alpha$, showing particle sizes $a$ (dark blue), $1.4a$ (light blue), $5a$ (grey) and $7a$ (white).
        (b) and (c) Constitutive flow curves showing the effect of adding grains (b) or colloids (c) to a colloidal (b) or granular (c) suspension initially at $\phi=0.55$.
        Particle addition is done holding both the volume of the initial species, $V_c$ (b) or $V_g$ (c), and the liquid volume $V_l$ fixed, so that both $\phi$ and $\alpha$ vary as detailed below.
        Vertical lines at selected $\pec$ highlight the variation of $\eta(\phi)$ at selected shear rates replotted in (d) when adding grains (blue) and colloids (orange).
        Volume ratios $\alpha$ and volume fractions $\phi$  are:
\protect\markerzero ~$\alpha=0,\phi = 0.55$; \protect\markerfive ~$\alpha=0.1, \phi = 0.576$; \protect\markerfour ~$\alpha=0.2,\phi = 0.604$; \protect\markerthree ~$\alpha=0.3,\phi = 0.636$; \protect\markertwo ~$\alpha=0.4,\phi = 0.671$; \protect\markerone ~$\alpha=0.5,\phi = 0.710$; \protect\markerseven ~$\alpha=0.6,\phi = 0.671$; \protect\markerten ~$\alpha=0.7,\phi = 0.636$; \protect\markereleven ~$\alpha=0.8,\phi = 0.604$; \protect\markertwelve ~$\alpha=0.9,\phi = 0.576$; \protect\markersix ~$\alpha=1,\phi = 0.55$.
}
\label{fig1}
\end{figure*}
%===========================
%===========================
%===========================
%===========================
%===========================

We first present two sets of constitutive curves in which
particles of one species are systematically added to a suspension initially comprising only the other, Fig.~\ref{fig1}(b) and (c).
Here we plot the suspension viscosity $\eta$ against $\pec$, with corresponding values of $\peg$ along the top axis, 
so that by reading vertically upward in Figs.~\ref{fig1}(b) and (c)
one isolates the affect of adding colloids or grains
at a fixed processing condition.

Starting from a pure colloidal system at
$\phi = 0.55$ and $\alpha=0$ (green circles, Fig.\ref{fig1}(b)), the rheology $\eta(\pec)$ is consistent with
earlier works~\cite{li2024simulating,mari2015discontinuous},
with a shear thinning regime at small $\pec\lesssim1$, followed by shear thickening to a frictional, rate-independent regime at large $\pec\gtrsim10^3$.
Systematically introducing grains by increasing $V_g$ while keeping both $V_c$ and $V_l$ constant,
 we find an increase in $\eta$ with $\alpha$ across all $\pec$. This is as might be expected, since the addition of grains in this manner also increases the total volume fraction $\phi$. Additionally, the shear thickening onset decreases monotonically to lower $\pec$ with increasing $\alpha$, accompanied by a steeper increase of $\eta$ above the thickening onset.
This mirrors experimental results demonstrating enhanced thickening when adding larger non-Brownian grains to suspensions of smaller colloids~\cite{cwalina2016rheology,madraki2017enhancing,madraki2018transition}.

Repeating this process, now adding colloids to an initially granular system at $\phi=0.55$ and $\alpha=1$,
produces a markedly different result. The rheology of the pure granular system (red circles, Fig.\ref{fig1}(c)) is similar to the pure colloidal system, but controlled by the granular P\'eclet number $\peg$.
However, the effect of adding colloidal particles by increasing $V_c$ and now keeping both $V_g$ and $V_l$ constant, is $\pec$-dependent. At low shear rates ($\pec \lesssim 0.1$), the viscosity $\eta$ increases monotonically with $\phi$, similar to our results adding grains to colloidal suspensions. 
However at higher $\pec$, adding colloids can actually {\it decrease} the suspension viscosity, despite the increase in $\phi$. This viscosity reduction is most pronounced around $\pec \sim 1$ ($\peg\sim10^2)$, where the viscosity decreases by up to a factor of $\sim7$ at $\alpha=0.7$ and $\phi=0.636$ before increasing and eventually jamming at higher $\phi$. At this shear rate, the pure granular suspension is in the high-viscosity frictional flow regime while the pure colloids are close their viscosity minimum. This nonmonotonic behavior persists at higher flow rates, where both the granular and colloidal suspensions have reached the frictional flow regime, though the magnitude of the viscosity reduction is notably reduced. 
While there are numerous experimental reports of bidisperse suspensions having lower viscosity than their monodisperse counterpart at the same $\phi$~\cite{chong1971rheology,gondret1997dynamic,zaman2006viscosity,guy2020testing}, our simulation results are the first to show a significant viscosity reduction at higher $\phi$ with the addition of colloidal solids. 
We further find that the range of shear rates over which shear thickening occurs is nonmonotonic in $\alpha$,
being narrower at $\alpha=1$ and $0.5$ compared to at intermediate values. 

Re-plotting the data along the vertical lines of Fig.~\ref{fig1}(b) and (c), showing $\eta(\phi)$ at fixed $\pec$, Fig.~\ref{fig1}(d), highlights the differing response to the addition of either colloids or grains.
(Note that $\alpha$ varies along each of these lines, as indicated by the symbols which match those in (b) and (c).)
The addition of grains to colloids (blue lines) leads to a monotonically increasing $\eta(\phi)$, with the most rapid increase at intermediate $\pec$.
In contrast, when adding colloids to grains (orange lines) $\eta(\phi)$ is nonmonotonic for $\pec\gtrsim0.1$, with the viscosity first decreasing as $\phi$ increases before eventually increasing again as the suspension approaches the jamming point. This viscosity reduction persists over a wide range of volume fraction, and is most notable for $\pec=1$ but also present for $\pec=10^3$, indicating that thermal motion enhances this effect but is not crucial.

\begin{figure*}
        \includegraphics[trim = 0.0cm 160mm 0cm 0cm, clip,width=\textwidth,page={2}]{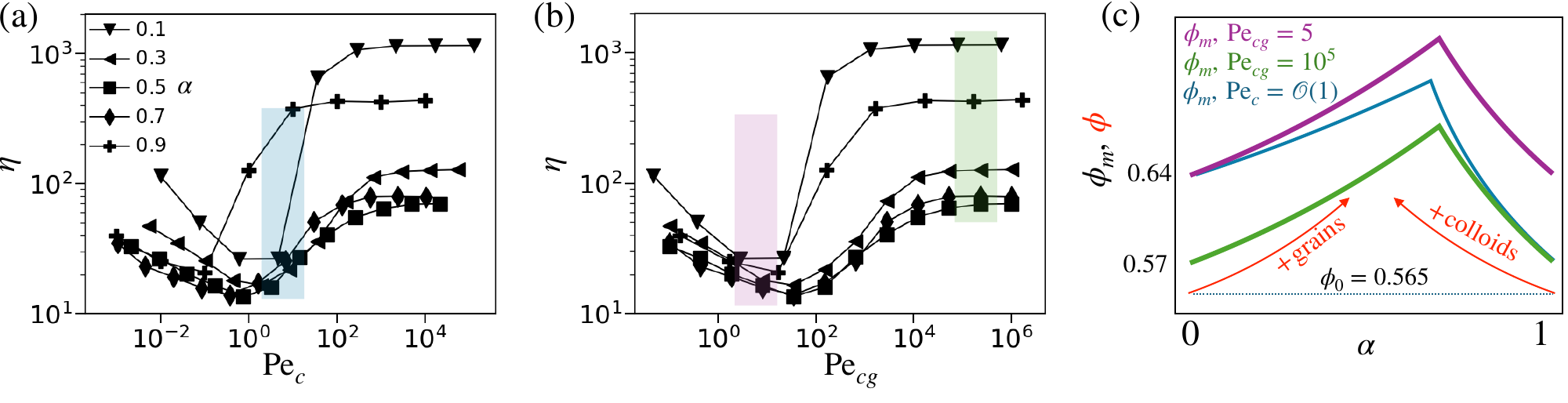}
        % includegraphics[width=\textwidth]{FIG2_jr.pdf}
        \caption{
        Mapping between P\'eclet number definitions.
        Constitutive flow curves at various $\alpha$ with fixed $\phi = 0.565$,
        reported as functions of (a) the colloidal $\pec$ and (b) the averaged $\pecg$.
        %In the latter case the curves each have minima at approximately the same $\pecg$;
        Coloured panels in (a) and (b) represent intermediate (blue), frictionless (purple) and frictional (green) states.
        (c) Sketch of the putative dependence of $\phi_m$ on $\alpha$ at the viscosity minimum where $\pecg=5$ (purple)
        and in the granular limit where $\pecg=10^5$ (green).
        Blue line interpolates between these as colloids are added at fixed $\pec$, so that the effective jamming point $\phi_m$ moves between green and purple lines as $\alpha$ decreases from 1 to 0.
        Red lines sketch the change in $\phi$ when adding either grains or colloids to a suspension at initial volume fraction $\phi_0$, highlighting the asymmetry in the change in the distance to jamming.
        Dotted blue line in (c) shows the volume fraction used in panels (a) and (b).
        }
        \label{fig2}
    \end{figure*}

To rationalize the above results in terms of changes in the critical jamming point $\phi_m$, which should now depend on both $\alpha$ and the shear rate, 
we must first reconsider how to define an appropriate P\'eclet number for these colloid/granular mixtures. 
Defining the P\'eclet number in terms of a single particle size (either $\pec$ or $\peg$)  allows for a systematic comparison of the constitutive curves at a fixed absolute shear rate $\dot\gamma$, but it becomes increasingly poor as a measure of the competition between diffusion and convection in mixtures at intermediate $\alpha$. This difficulty can be seen in constitutive flow curves $\eta(\pec)$ at fixed $\phi$ and varying $\alpha$, where the transition from shear thinning to shear thickening shifts by over two orders of magnitude in $\pec$, Fig.~\ref{fig2}(a).

%===========================
%===========================
%===========================
%===========================
%===========================
%===========================
%===========================
%===========================
%===========================
%===========================

We empirically find that plotting viscosity against an averaged P\'eclet number $\pecg={6 \pi \eta \dot\gamma a_{cg}^3}/{k_B T}$, defined using an $\alpha$-weighted mean particle size $a_{cg}=(1-\alpha)a_c + \alpha a_g$, roughly aligns these flow curves,  Fig.~\ref{fig2}(b). The $\eta(\pecg)$ flow curves now all have minima over a narrow range of $\pecg\sim10$, and reasonable overlap in the shear thinning regime at lower $\pecg$. At higher $\pecg$, the shear thickening dynamics show a stronger $\alpha$-dependence. Suspensions closer to the monodisperse limit (either $\alpha\to0$ or $\alpha\to1$) have both an increased frictional (high-shear) viscosity and a more abrupt shear thickening transition ($10\lesssim\pecg\lesssim10^3$), while more evenly mixed suspensions ($\alpha\approx0.5$) exhibit a broader thickening transition ($10\lesssim\pecg\lesssim10^4$) to a lower high-shear viscosity despite all having the same $\phi$.

The aligned shear-thickening flow curves in Fig.~\ref{fig2}(b) indicate that we can regard the intermediate (pink) and high $\pecg$ (green) states as frictionless and frictional respectively. Drawing from prior studies of binary sphere packings~\cite{srivastava2021jammingbi, singh2024rheology}, we sketch $\phi_m(\alpha)$ in these two limits in Fig.~\ref{fig2}(c). We compute these curves using an analytic packing model~\cite{yu1991estimation}, but note that the key features of these curves are insensitive to the specific choice of model~\cite{anzivino2023estimating,roquier2024centgran}. The limits for the monodisperse cases at $\alpha=(0,1)$ are set to $\phi_m^{\mu=0}=0.64$ and $\phi_m^{\mu=1}=0.57$. In binary mixtures, $\phi_m(\alpha)$ is nonmonotonic and asymmetric, with a maximum near $\alpha \approx 0.7$ in both the frictional and frictionless cases. Due to this asymmetry, both the frictional and frictionless $\phi_m$ curves increase more rapidly when $\alpha$ decreases from an initially granular system ($\alpha=1$) compared to increasing $\alpha$ from an initially colloidal system ($\alpha=0$).

Varying $\alpha$ at fixed $\phi=\phi_0$ in either the intermediate or high-$\pecg$ regimes alters the ratio $\phi/\phi_m$ between our fixed packing fraction (horizontal dotted line in Fig.~\ref{fig2}(c)) and the relevant $\alpha$-dependent $\phi_m$. Close to jamming, we expect that reducing this ratio reduces the suspension viscosity (Eq.~\ref{eq:visc_scale}). This is consistent with our viscosity measurements in these two highlighted flow regimes, with the viscosity reduction at intermediate $\alpha$ most pronounced in frictional high-$\pecg$ regime due to the closer proximity to jamming.

If we vary $\alpha$ at fixed $\phi$ now at fixed intermediate $\pec=\mathcal{O}(1)$ within the shear thickening transition,
the relevant jamming point shifts between the two branches as we vary $\alpha$, with $\phi_m\to\phi_m^{\mu=0}$ in the colloidal ($\alpha\to0$) limit and $\phi_m\to\phi_m^{\mu=1}$ in the granular ($\alpha\to1$) limit. This is sketched schematically by the solid blue line in Fig.~\ref{fig2}(c). As in the fixed $\pecg$ examples, the viscosity again varies nonmonotonically with $\alpha$ but is now notably higher in the granular limit, reflecting the asymmetry in $\phi_m(\alpha)$.

Adding particles of a different species to an initially monodisperse suspension (fixing both the liquid volume and volume of the initial species, as in Fig.~\ref{fig1}) increases $\phi$ [red arrows in Fig.~\ref{fig2}(c)].
Comparing the sketched $\phi(\alpha)$ with the respective $\phi_m$ lines,
 adding grains moves $\phi$ closer to $\phi_m$ for any choice of $\pe$, whereas adding colloids initially moves $\phi$ further from $\phi_m$.
This holds even in the large-$\pecg$ limit, implying an initial viscosity reduction when adding smaller particles to suspension of larger ones even when both species are non-Brownian (and fully frictional). At intermediate $\pec=\mathcal{O}(1)$,  the shift in $\phi_m$ between frictional and frictionless jamming points marked by the blue line in Fig.~\ref{fig2}(c) suggests an enhanced viscosity reduction when adding colloids to a granular suspension near the colloidal-to-granular transition.

%===========================
%===========================
%===========================
%===========================
%===========================
\begin{figure*}
    \centering
    \includegraphics[trim = 0cm 165mm 0mm 0cm, clip,width=1\textwidth,page={3}]{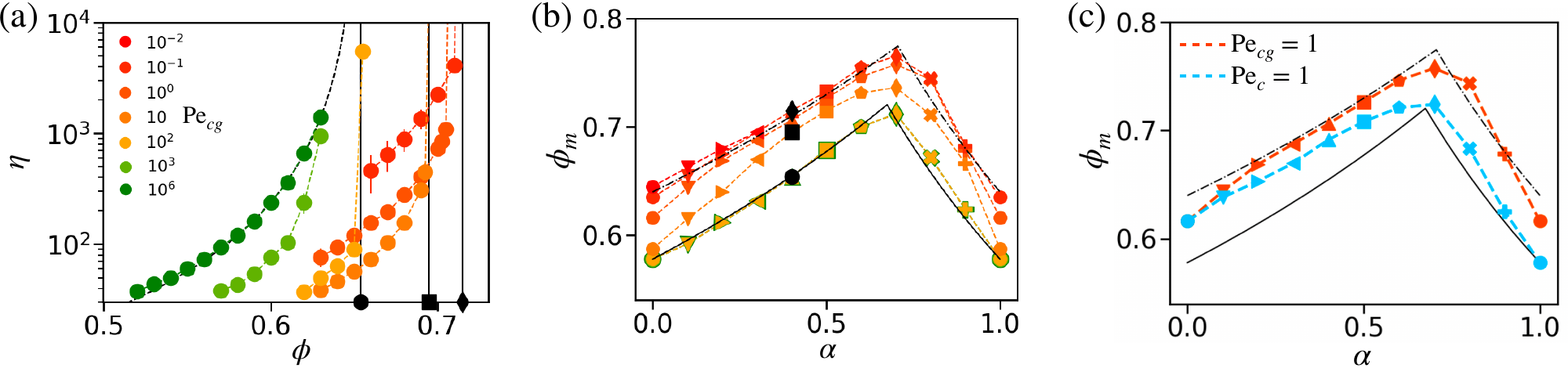}
    \caption{
    Jamming volume fraction at the colloidal-to-granular transition.
    (a) Viscosity divergence with $\phi$ for $\alpha = 0.4$, at a range of $\pecg$.
    Data at $\pecg = 10^6$ are fit to $\eta(\phi) \sim (1-\phi/\phi_m)^{-\beta}$, shown by the black dashed line.
    Three representative $\phi_m$ for $\pecg=100,10,0.1$ are marked left-to-right by black vertical lines, and correspond to the black points in (b).
    (b) $\phi_m$ measured across a range of $\pecg$ and $\alpha$, with the former indicated by the same color legend as (a), and the latter with the same marker shapes as in Fig.~\ref{fig1}.
    The solid and dashed black lines represent model predictions~\cite{yu1991estimation} and provide geometric $\phi_m$ estimates taking the values at $\alpha=(0,1)$ as inputs;
    (c) Example of $\phi_m(\alpha)$ at fixed $\pec=1$, obtained by interpolating through the fixed $\pecg$ data in (b). Shown for comparison is $\phi_m(\alpha)$ measured at fixed $\pecg=1$.
    }
    \label{fig3}
\end{figure*}
%===========================
%===========================
%===========================
%===========================
%===========================

\section{RESULTS: Mapping the jamming point}
We now test this schematic picture using our simulation model.
To do so, we first map the jamming point $\phi_m(\alpha, \pecg)$ over a wide range of our mixed P\'eclet number $10^{-2} \leq \pecg \leq 10^6$. This is done from steady-state viscosity measurements on systems spanning a range of $\alpha$ with narrow increments of $\phi$ to clearly resolve the viscosity divergence. Example results for $\alpha = 0.4$ are shown in Fig.~\ref{fig3}(a).
At large $\pecg$ the viscosity follows a power-law divergence as $\phi\to\phi_m$ (Eq.~\ref{eq:visc_scale}, black dashed line), as expected for an athermal suspension.
Reducing the P\'eclet number $\pecg \leq 10^3$ changes the form of $\eta(\phi)$, with the smooth power-law divergence replaced by an abrupt viscosity jump. We identify $\phi_m$ at these higher $\pecg$ from the maximum of the gradient $\partial \eta/\partial \phi$ (black vertical lines in Fig.~\ref{fig3}(a) and corresponding black points in Fig.~\ref{fig3}(b)), as the behavior of $\eta(\phi)$ below this point gives no indication of imminent jamming. This change in behavior can be seen clearly in the data for $\pecg=10^2$ in Fig.~\ref{fig3}(a), where the viscosity is dramatically reduced relative to that at $\pecg=10^6$ though there is no discernible change in the jamming point at $\phi_m=\phi_m^{\mu=1}$. This suggests that the weak Brownian motion is sufficient to decrease the contact stress but not to significantly change the jamming point.

Further reducing $10^{-2} < \pecg \leq 10^{2}$, the jamming point $\phi_m$ shifts upwards approaching the frictionless limit $\phi_m^{\mu=0}$, suggesting that within this range of P\'eclet number Brownian motion is now sufficient to mobilize and release frictional contacts and thus shift the jamming point. At small $\pecg\lesssim10$,
the viscosity curves shift vertically upward as the shear rate is reduced,
reflecting an increasing contribution from the Brownian stress.

The full $\phi_m(\alpha)$ curves at fixed values of $\pecg$ shown in Fig.~\ref{fig3}(b) validate the schematic picture sketched in Fig.~\ref{fig2}(c), with the high-shear ($\pecg\geq10^2$) and low-shear ($\pecg\sim10^{-2}$) branches matching our estimates based on the monodisperse frictional and frictionless jamming points. At intermediate $\pecg$, $\phi_m(\alpha)$ falls between these two limiting branches, with symmetric monodisperse limits $\phi_m(\alpha=0, \pecg)=\phi_m(\alpha=1, \pecg)$. Here $\pecg$ is constant along each plotted line,
so that the effective roles of Brownian versus convective transport are
constant despite changes in composition. The transition between these two limiting branches gives the $\pecg$-controlled shear thickening shown in Fig.~\ref{fig2}(b). This increase in $\phi_m$ with decreasing $\pecg$ mirrors the shift from frictional to frictionless $\phi_m$
in bidisperse non-Brownian suspensions with a similar size ratio~\cite{singh2024rheology}, indicating that thermal motion here plays a key role freeing frictional constraints. 

We next map these jamming curves back into $\pec$ space to describe flows where the absolute shear rate is fixed, independent of composition. 
To obtain $\phi_m$ for fixed $\pec$ and $\alpha$,  
we first compute the corresponding $\pecg$ and then use nearest-neighbor linear interpolation with the measured $\phi_m(\alpha,\pecg)$ results shown in Fig.~\ref{fig3}(b) to estimate $\phi_m(\alpha,\pec)$.
As initially sketched in Fig.~\ref{fig2}(c), these $\phi_m(\alpha)$ profiles at fixed $\pec$ are no longer constrained to be symmetric in the two monodisperse limits. 
Plotting $\phi_m(\alpha)$ at fixed $\pec=1$, Fig.~\ref{fig3}(c), the jamming point coincides with the value at $\pecg=1$ at $\alpha=0$ (by definition), closer to the frictionless branch, but peels below this curve as $\alpha$ increases and transitions to the frictional branch in the granular ($\alpha\to1$) limit.
Finally, with these tools to compute $\phi_m(\alpha)$ at fixed $\pec$, we now have a comprehensive framework to understand the differing effects of adding either colloids or grains to an initially monodisperse suspension of the other species. 

\begin{figure*}
    \centering
    \includegraphics[trim = 0cm 12mm 0mm 0cm, clip,width=0.95\textwidth,page={4}]{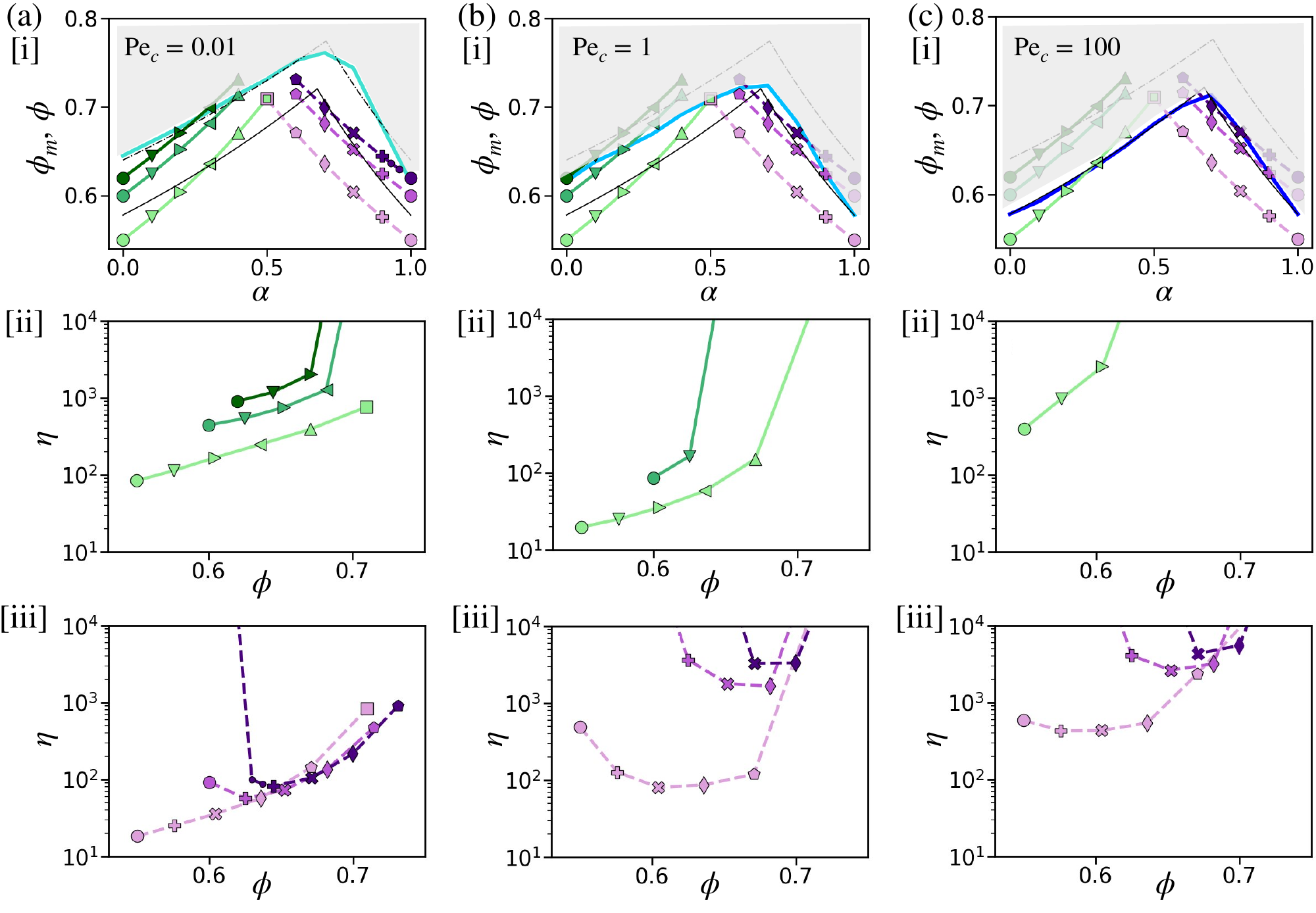}
    \caption{
    Effect of particle addition on viscosity at three fixed values of $\pec$.
    Shown in (a)[i], (b)[i] and (c)[i] are theoretical predictions (black lines) together with interpolated plots of $\phi_m(\alpha)$ at $\pec=0.01$, $1$ and $100$ (solid blue lines).
    Solid green and purple markers in [i] are not jamming points but represent changes in $\phi$ under particle addition.
    Shaded regions show parameter values for $\alpha$ and $\phi$ for which the system is jammed.
    For each value of $\pec$ we explore six cases of particle addition, adding either grains (green) or colloids (purple) to suspensions initially at $\alpha=0$ and $\alpha=1$ respectively, initially with $\phi=0.55$, $0.6$ and $0.62$ (shown by light  to dark lines and markers coloured green (grain addition) and purple (colloid addition).
    Panel rows [ii] and [iii] show the consequent variations in the viscosity $\eta$ with volume fraction $\phi$ for addition of grains [ii] and colloids [iii].
    }
    \label{fig4}
\end{figure*}

We show interpolated plots of $\phi_m(\alpha)$ at fixed $\pec=0.01$, $1$ and $100$ in figures Fig.~\ref{fig4}(a)[i]-(c)[i] respectively, capturing cases where $\phi_m(\alpha)$ either remains along the upper frictionless branch [Fig.~\ref{fig4}(a)], remains along the lower frictional branch [Fig.~\ref{fig4}(c)] or transitions between the two as $\alpha$ is varied [Fig.~\ref{fig4}(b)]. The inaccessible jammed region above these curves is shaded in grey. Solid symbols in the same panels show the volume fraction $\phi$ when either adding grains to a colloidal suspension (green) or colloids to a granular suspension (purple) starting from differing initial volume fractions (bottom to top $\phi=0.55$, $0.6$, $0.62$). As in Fig.~\ref{fig1}, this particle addition is done holding both the liquid volume $V_l$ and the volume of the initial species (either $V_c$ or $V_g$) fixed.

%===========================
%===========================
%===========================
%===========================
%===========================
%===========================
%===========================
%===========================
%===========================
%===========================

Adding grains to an initially colloidal ($\alpha=0)$ suspension moves $\phi(\alpha)$ closer to the jamming point $\phi_m(\alpha)$ independent of $\pec$ or the initial $\phi$, so that the corresponding $\eta(\phi)$ curves always increases monotonically, Fig.~\ref{fig4}(a)[ii]-(c)[ii], eventually diverging as they cross the $\phi_m$ boundary. This accounts for the monotonic rise in $\eta$ shown in Fig.~\ref{fig1}(a). Interestingly, the measured viscosities $\eta(\phi)$ below jamming are lower for $\pec=1$ than at $\pec=0.01$, despite the former being closer to jamming.
This reflects the decreasing contribution of the Brownian stress at higher $\pec$,
 also seen in Fig.~\ref{fig3}(a).
As $\pec$ increases, the window of accessible volume fractions shrinks as particle friction becomes more important,
so that in Fig.~\ref{fig4}(c) the systems starting at $\alpha=0$ and $\phi=0.6$, $0.62$ are already above their respective jamming points.
Thus the scope for manipulating the viscosity of initially colloidal suspensions by adding grains is reduced with increasing $\pe$ as the jamming point shifts to lower volume fractions.

In contrast, the effect of adding smaller colloids to an initially granular ($\alpha=1)$ suspension varies with $\pec$. The asymmetry in the $\phi_m(\alpha)$ curves, with maxima around $\alpha\sim 0.6$ independent of $\pec$, results in $\phi$ increasing with decreasing $\alpha$ slower than the growth of $\phi_m$. In other words, adding colloids to a granular suspension can move the system \emph{out} of the inaccessible region, allowing the jammed suspension to flow despite the \emph{increase} in solids content. This flowable region extends until the $\phi_m(\alpha)$ boundary is again crossed at some point below the maxima in the $\phi_m(\alpha)$ curve. 
This suggests in principle that adding colloids while keeping $0.6<\alpha<1$ should always move the suspension further away from jamming. This holds at $\pec=1$ and $\pec=100$, where in both cases it is possible to fluidise initially jammed suspensions by adding colloidal particles, Fig.~\ref{fig4}(b)[iii]-(c)[iii], going from a divergent to finite viscosity with increasing $\phi$.  
Starting from a granular suspension below jamming ($\phi = 0.55$), the addition of colloids reduces the suspension viscosity, with this reduction most pronounced around $\pec=1$ where $\phi_m(\alpha)$ rises most steeply with decreasing $\alpha$ and then falls more gradually below the maximum. 

At $\pec = 0.01$, where Brownian contributions to the viscosity are more significant, the effect of adding colloids is more complicated. We do not observe a viscosity reduction when adding colloids to a granular suspension at a relatively low initial packing density of $\phi=0.55$ [Fig.~\ref{fig4}(a)[iii], see also the dot-dashed orange line in Fig.\ref{fig1}(d)]. In this case, given our greater initial distance from jamming,  Eq.~\ref{eq:visc_scale} likely requires higher-order terms to fully describe $\eta(\phi)$, so that we cannot simply map between the viscosity and the distance to jamming. If we instead consider initial packing densities either just above or very close to jamming ($\phi=0.62$ and $\phi=0.6$), we recover the un-jamming effect and initial viscosity reduction with the addition of colloids, demonstrating that this framework for understanding the effect of added colloids holds even at low $\pec$ provided we are sufficiently close to jamming.

Ultimately our modeling predicts that it is the asymmetry in the jamming plots that allows for the contrasting effect of adding colloids or grains.
Adding grains always moves $\phi$ closer to $\phi_m$;
adding colloids can move $\phi$ further from $\phi_m$,
exaggerated when considering fixed $\pec$ so that $\phi_m$
is both $\pe$- and $\alpha$-dependent. Interestingly, we can capture this phenomenology considering only the frictional and frictionless limits of $\phi_m$, largely ignoring the colloidal glass transition at some $\phi_G< \phi_m^{\mu=0}$ even at our lowest $\pec=0.01$. While we would expect this glass transition to play a critical role in the $\pec\to0$ limit, our results indicate that the dominant role of thermal motion in the range of finite shear rates explored here is simply to inhibit frictional particle contacts.

\section{\label{sec:level4}Conclusion}
We have systematically explored the constitutive curves of suspensions comprising mixed colloids and grains,
demonstrating the complex dependence on the composition $\alpha$ and the shear rate quantified through $\pec$ and $\pecg$.
As well as the expected shear thinning and thickening behavior, the simulation predicts a counterintuitive viscosity drop upon addition of colloids to an initially pure granular suspension,
an effect that is most pronounced at shear rates near the colloidal-to-granular crossover.
We rationalized all of the observed flow curves
by mapping the jamming volume fraction
as a function of $\alpha$ and $\pe$.
By introducing a rescaled P\'eclet number, we quantified the interplay between shear flow and thermal motion, providing a clear separation of
granular, colloidal and crossover regimes for bidisperse Brownian suspensions.
Additionally, we identified that $\phi_m$ increases as thermal energy increases, which aligns with the shift from frictional to frictionless behavior.

These findings provide important insights for the development and optimization of materials where viscosity control is critical. The ability to predict and manipulate the viscosity of bidisperse suspensions by adjusting $\alpha$, $\dot\gamma$, and $k_BT$ has potential applications in industries such as concrete formulation~\cite{roussel2010steady}, dip coating~\cite{jeong2022dip}, and printing~\cite{thievenaz2021droplet}. Future research could focus on extending this framework to more complex particle shapes and interactions, offering broader applicability to real-world systems.

\vspace{3mm}
\begin{acknowledgments}
C.N.~acknowledges support from the Royal Academy of Engineering under the Research Fellowship scheme and from the Leverhulme Trust under Research Project Grant RPG-2022-095.
Scripts required to reproduce the data reported in this paper are available on request.
For the purpose of open access, the authors have applied a Creative Commons Attribution (CC BY) licence to any Author Accepted Manuscript version arising from this submission.
Declaration of Interests: the authors report no conflict of interest.
\end{acknowledgments}

\bibliographystyle{apsrev4-1}
\bibliography{main}

%\clearpage
\appendix
\label{appendix}
\section{Simulation details}
A comprehensive description of our model is given in~\cite{li2024simulating}.
In short,
particle motion is governed by a set of Langevin equations incorporating forces $\mathbf{F}$ originating from direct particle contacts, hydrodynamics, and Brownian motion. The translational equation of motion for particle $i$ with position $\mathbf{x}_i$ and mass $m_i$ is
\begin{equation}
m_i \frac{d^2\mathbf{x}_i}{dt^2} = \sum_j \mathbf{F}_{i,j}^{C} + \mathbf{F}_{i}^{H,D} + \sum_j \mathbf{F}_{i,j}^{H,L} + \mathbf{F}_{i}^{B,D} + \sum_j \mathbf{F}_{i,j}^{B,L}\text{.}
\end{equation}
Forces with subscript $i$ are single-body whereas those with subscript $i,j$ act pairwise between particles $i$ and $j$.
Superscripts C, H, B, D and L refer
to forces arising due to
contacts (C),
hydrodynamics (H)
and Brownian (B) effects,
with the latter two having
drag (D) and
lubrication (L) terms.
An equivalent set of torques are computed and used to update the rotational velocities of the particles,
but these are omitted here for brevity.

Contact forces are modeled as linear springs,
with the normal force between $i$ and $j$ proportional to their overlap $\delta_{i,j}$:
\begin{equation}
\mathbf{F}_{i,j}^{C} = k_n \delta_{i,j} \mathbf{n}_{i,j} - k_t \boldsymbol{\xi}_{i,j}\text{,}
\end{equation}
with $k_n$ and $k_t$ the spring constants and
$\mathbf{n}_{i,j}$ the unit vector pointing from $i$ to $j$.
The tangential force is linear in $\boldsymbol{\xi}_{i,j}$,
the accumulated displacement of the particle
pair perpendicular to $\mathbf{n}_{i,j}$ since the initiation of the contact.
Tangential forces are regulated by a Coulombic friction coefficent set as $\mu=1$ throughout.
The contact stress tensor $\mathbb{\Sigma}^{C}$ is obtained by summing the outer product $-\mathbf{F}_{i,j}^{C}\otimes \mathbf{r}_{i,j}$
(with $\mathbf{r}_{i,j}$ the center-to-center vector)
over all contacting pairs.

Hydrodynamic interactions comprise Stokes drag and pairwise lubrication.
The former is given for particle $i$ as:
\begin{equation}
\mathbf{F}_{i}^{H,D} = 6 \pi \eta a_i (\mathbf{U}_{\infty}(\mathbf{x}_i) - \mathbf{U}_i)\text{,}
\end{equation}
and the latter are given for $i$ and $j$ as:
\begin{equation}
\begin{split}
\mathbf{F}_{i,j}^{H,L} = &(X_{11}^{A} \mathbb{N}_{i,j} + Y_{11}^{A}\mathbb{ T}_{i,j})(\mathbf{U}_j - \mathbf{U}_i) \\
 &+ Y_{11}^{B}(\boldsymbol{\Omega}_i \times \mathbf{n}_{i,j}) + Y_{21}^{B}(\boldsymbol{\Omega}_j \times \mathbf{n}_{i,j})\text{.}
\end{split}
\end{equation}
Here $\mathbb{N}$ and $\mathbb{T}$ are the normal and tangential projection operators,
$\mathbf{U}_i$ is the velocity of particle $i$,
$\mathbf{U}_{\infty}(\mathbf{x}_i) $ is the liquid streaming velocity at the position of particle $i$
and $\boldsymbol{\Omega}_i$ is the
rotational velocity of particle $i$.
The scalar prefactors $X$ and $Y$ encode the geometry of the interacting pair
and are given by~\cite{jeffrey1992calculation}. 
The hydrodynamic stress tensor $\mathbb{\Sigma}^{H}$ has contributions from drag and lubrication,
the latter obtained by summing $-\mathbf{F}_{i,j}^{H}\otimes \mathbf{r}_{i,j}$
over all interacting pairs.

Brownian forces are generated to satisfy fluctuation-dissipation theorem,
with the single body and pairwise forces given respectively as:
\begin{equation}
\mathbf{F}_{i}^{B,D} = \sqrt{\frac{2k_b T}{\Delta t}} \sqrt{6 \pi \eta a_i} \boldsymbol{\psi}_i\text{,}
\end{equation}
\begin{equation}
\mathbf{F}_{i,j}^{B,L} = \sqrt{\frac{2k_b T}{\Delta t}} \left(\sqrt{X_{11}^{A}} \mathbb{N}_{i,j} + \sqrt{Y_{11}^{A}} \mathbb{T}_{i,j} \right) \boldsymbol{\theta}_{i,j}\text{,}
\end{equation}
where $k_b T$ is the thermal energy,
$\Delta t$ is the time step and
$\boldsymbol{\psi}_i$, $\boldsymbol{\theta}_{i,j}$
are unit vectors of random direction.

The parameters used to generate the data in Fig.~\ref{fig1} are as follows,
with the dimensions of each quantity given in terms of
mass ($M$),
length ($L$),
and time ($T$),
and (--) representing dimensionless quantities.
Small particles have
$a_S = 1$  $(L)$,
$k_n = 10{,}000$ $(M/T^2)$,
$k_t = 7{,}000 $ $(M/T^2)$,
and density $\rho = 0.1$ $(M/L^3)$.
Large particles have
$a_L = 5$ $(L)$,
$k_n = 50{,}000 $ $ (M/T^2)$,
$k_t = 35{,}000 $ $ (M/T^2)$,
and $\rho = 0.004 $ $ (M/L^3)$.
The liquid viscosity is $\eta = 0.1$ $(M/L T)$
and the time step is $\Delta t = 0.0001$ $(T)$.
These parameters ensure approximately inertia-free and hard sphere conditions as described in~\cite{li2024simulating}.
The values of $\dot\gamma$ and $k_BT$ used to achieve each desired P\'eclet are given in Table~\ref{table1}.
\begin{table}[h]
    \caption{Parameters used to generate the P\'eclet numbers in Figs.~\ref{fig1}(a) and (b), and their dimensions.}
    \begin{tabular}{l|p{65pt} p{50pt} p{50pt}}
    \hhline{====}
        {Quantity:}
        &Small particle P\'eclet number
        &Shear rate
        &Thermal energy \\
        &$\pec$
        &$\dot{\gamma}$
        &$k_BT$ \\
        \hline
        {Dimensions:} & -- & $1/T$ & $ML^2/T^2$
        \\
        \hline 
         Fig.~\ref{fig1}(b):& $1.0\times10^{-2}$ & $4.5\times10^{-4}$ & $8.5\times10^{-2}$ \\
         & $7.7\times10^{-2}$ & $3.5\times10^{-3}$& $8.5\times10^{-2}$ \\
         & $6.0\times10^{-1}$ & $2.7\times10^{-2}$& $8.5\times10^{-2}$ \\
         & $4.6\times10^{0}$ & $4.5\times10^{-2}$& $1.8\times10^{-2}$\\
         &$3.6\times10^{1}$  & $4.5\times10^{-2}$& $2.4\times10^{-3}$\\
         & $2.8\times10^2$  & $4.5\times10^{-2}$& $3.0\times10^{-4}$\\
         & $2.2\times10^3$  & $4.5\times10^{-2}$& $3.9\times10^{-5}$\\
         & $1.7\times10^4$  & $4.5\times10^{-2}$& $5.1\times10^{-6}$\\
         & $1.3\times10^5$  & $4.5\times10^{-2}$& $6.6\times10^{-7}$\\
         & $1.0\times10^6$  & $4.5\times10^{-2}$&$8.5\times10^{-8}$\\
        \hline  
         Fig.~\ref{fig1}(c):& $1.0\times10^{-3}$ & $4.5\times10^{-5}$ & $8.5\times10^{-2}$ \\
         & $1.0\times10^{-2}$ & $4.5\times10^{-4}$& $8.5\times10^{-2}$ \\
         & $1.0\times10^{-1}$  & $4.5\times10^{-3}$& $8.5\times10^{-2}$ \\
         & $1.0\times10^{0}$  & $4.5\times10^{-2}$& $8.5\times10^{-2}$\\
         & $1.0\times10^{1}$ & $4.5\times10^{-2}$& $8.5\times10^{-3}$\\
         & $1.0\times10^{2}$   & $4.5\times10^{-2}$& $8.5\times10^{-4}$\\
         & $1.0\times10^{3}$  & $4.5\times10^{-2}$& $8.5\times10^{-5}$\\
         & $1.0\times10^{4}$   & $4.5\times10^{-2}$& $8.5\times10^{-6}$\\
         & $1.0\times10^{5}$  & $4.5\times10^{-2}$& $8.5\times10^{-7}$\\
         &$1.0\times10^{6}$
         &$4.5\times10^{-2}$
         &$8.5\times10^{-8}$\\
        \hline 
        \hline
    \end{tabular}
    \label{table1}
\end{table}

\end{document}